\documentclass[9pt,twocolumn,twoside]{optica}
\setboolean{shortarticle}{true}
\setboolean{minireview}{false}
\usepackage[normalem]{ulem}

\title{Simultaneous laser-driven X-ray and two-photon fluorescence imaging of atomizing sprays}

\author[1,*]{D. Gu\'enot}
\author[1]{K. Svendsen}
\author[1]{J. Bj\"orklund Svensson}
\author[1]{H. Ekerfelt}
\author[1]{A. Persson}
\author[1]{O. Lundh}
\author[2]{E. Berrocal}

\affil[1]{Division of Atomic Physics, Department of Physics, Lund University, PO Box 118, SE-22100, Lund, Sweden.}
\affil[2]{Division of Combustion Physics, Department of Physics, Lund University, PO Box 118, SE-22100, Lund, Sweden.}
\affil[*]{Corresponding author: diego.guenot@fysik.lth.se}

\doi{\url{https://doi.org/10.1364/OPTICA.378063}}

\begin{abstract}

In this letter we report for the first time the possibility of visualizing an atomizing spray by simultaneously recording X-ray absorption and 2-photon laser-induced fluorescence imaging. This unique illumination/detection scheme was made possible thanks to the use of soft X-rays emitted from a laser-driven X-ray source. An 800\,mJ laser pulse of 38\,fs duration is used to generate an X-ray beam with up to $4\times10^8$ photons ranging from 1 to 10\,keV, allowing projection radiography of water jets generated by an automotive port fuel injector. 
In addition, a fraction of the laser pulse ($\sim$ 10\,mJ) is employed to form a light sheet and to induce 2-photon fluorescence in a dye added to the water. The resulting high-contrast fluorescence images provide fine details of the spray structure, with reduced blur from multiple light scattering, while the integrated liquid mass is extracted from the X-ray radiography.
In this proof-of-principle we show that the combination of these two highly complementary techniques, both in the visible and in the soft X-ray regime, is very promising for the future characterization of challenging spray, as well as for further understanding the physics of liquid atomization.

\end{abstract}

\setboolean{displaycopyright}{true}

\begin{document}

\maketitle


[© 2019 Optical Society of America]. Users may use, reuse, and build upon the article, or use the article for text or data mining, so long as such uses are for non-commercial purposes and appropriate attribution is maintained. All other rights are reserved.

\begin{figure*}[t]
\centering
\fbox{\includegraphics[width=179mm,scale=1]{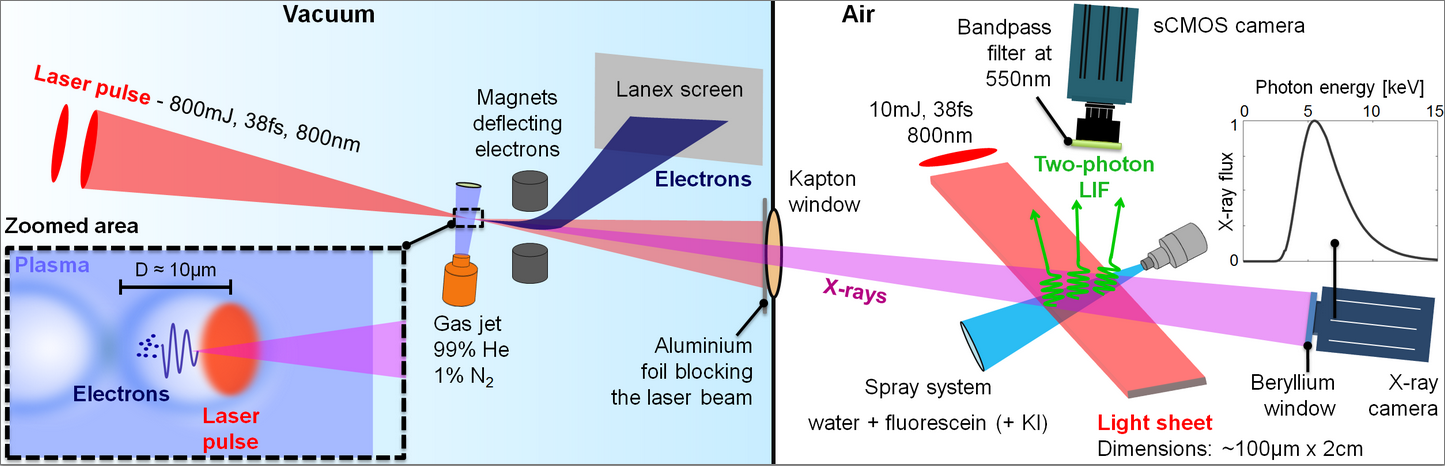}}
\caption{ Illustration of the experimental setup: An X-ray camera detects the transmitted X-rays while a sCMOS camera record simultaneously the fluorescence generated from a two-photon excitation process. The zoomed inset illustrates the laser plasma X-ray generation: The background plasma density is shown in blue, the laser pulse is in red and the X-ray beam is indicated in magenta while the detected spectrum is shown above the X-ray camera. A small part of the laser pulse is extracted before focusing (not shown here) then redirected towards the spray and focused into a light sheet. The blue curve represents a typical electron trajectory. The emitted electrons are deflected from the X-rays beam using a strong dipole magnet and are imaged on a Lanex screen to obtain the electron beam spectra. An Aluminium foil + Kapton vacuum window allows blocking the laser radiation while letting the X-ray beam exiting the vacuum chamber.}
\label{fig:setup}
\end{figure*}

Atomizing sprays are used for a variety of applications such as applying paint or chemicals for surface treatments, cutting material by means of water-jet cutters, cooling hot environments or surfaces, injecting ink for printers, treating crops in agriculture $etc$. Nonetheless, liquid jet atomization is most extensively used for combustion purposes such as in internal combustion engines (e.g. Gasoline Direct Injection (GDI) and Diesel engines) as well as in gas turbine aero-engines. In these cases, a precise amount of liquid fuel needs to be injected, disintegrated, evaporated and properly mixed prior to combustion in order to optimize the combustion efficiency. In addition, the use of alternative bio-fuels may require different injection strategies as the liquid properties, such as surface tension, liquid density and viscosity, can significantly differ from one fuel to another. Such changes in the liquid properties directly impact the atomization process and, thus the efficiency and resulting  emission of pollutants. 
\\
The use of imaging techniques for spray characterization is of utmost importance in order to: 1) Provide detailed information related to the process of spray formation. 2) Quantitatively describe the formed cloud of droplets (e.g. by measuring the droplet size, velocity vectors and liquid volume fraction).
However, the main challenge in visualizing an optically dense spray is to mitigate the effects of multiple light scattering from the surrounding droplets, blobs and other liquid bodies which are present outside of the image plane. This out-of-focus light contribution results in visibility reduction and image distortion. The efforts and means employed to overcome issues related to multiple light scattering in atomizing sprays have increased over the past two decades, leading to the development and application of a variety of advanced imaging techniques. 
\\
A first approach consists in selectively filtering out photons that have undergone multiple scattering events. This filtering process can be done by time-gating photons prior to detection, via transillumination - e.g. Ballistic Imaging \cite{Paciaroni2004,Linne2009,Idlahcen2012} - or back-scattering \cite{Rahm2016} detection. 
A second approach consists in suppressing the unwanted light intensity contribution after image recording. This is the case for Structured Illumination based techniques where a spatially modulated illumination is used to encode the incident light. The approach has been mainly employed  for light sheet imaging - e.g. Structured Laser Illumination Planar Imaging, SLIPI \cite{Berrocal2008,Kristensson2018,Berrocal2012} - but it can also be used for transillumination detection \cite{Berrocal2016} and be associated to computer tomography for 3D  reconstructions of the spray region \cite{Kristensson2012}.  
A third approach consists in directly reducing the generation of multiple light scattering. This can be done either by implementing two-photon laser-induced fluorescence (2p-LIF) \cite{Berrocal2019} or by using X-rays \cite{Heindel2019}.
\\
The advantage of 2p-LIF detection in optically dense sprays is that it provides much higher image contrast than for one-photon liquid LIF or for elastic Mie scattering as recently demonstrated in \cite{Berrocal2019}. The main reason for this is that multiply-scattered photons spread in space and time, greatly reducing the probability of having two photons simultaneously absorbed. On the contrary, at the location where the illuminating light sheet is focused the probability for the 2p-LIF process to occur is at its highest, providing a signal that is generated only at the object plane of the camera objective. Consequently, light sheet 2p-LIF provides images with high fidelity of the liquid bodies and the presence of voids, even in the spray formation region. Finally, the main advantage of 2p-LIF light sheet imaging over SLIPI is that it does not require the recording of several modulated sub-images (to preserve the image spatial resolution \cite{Mishra2017}).  
\\
Unlike visible light, the refractive index of the injected liquid, such as water, becomes close to unity for photons in the keV range, while the absorption cross-section is above $10^2\,\text{cm}^2/\text{g}$. Thus, in X-ray radiography the amount of scattered radiation is negligible in comparison with absorption making this approach the most reliable for measuring liquid mass distribution in the near-nozzle region, where large and irregular liquid structures are still present and where the liquid density is the highest. As the liquid mass distribution is related to the rate of liquid breakup and to gas entrainment, this quantity is critical for understanding how sprays are formed \cite{Linne2012}.\\
Due to the short time scale ($\approx 1$\,µs) of the breakup process and the high X-ray flux required for reaching such time resolution, most of the research efforts related to radiography of transient sprays have been accomplished at synchrotron facilities such as the Advanced Photon Source at Argonne National Laboratory. The technique has been successfully and extensively used since 2000 \cite{Powell2000} for various spray studies: From the observation of shockwaves generated by high-pressure diesel sprays \cite{MacPhee2002} to, more recently, the analysis of primary breakups using high-speed X-ray radiography \cite{Halls2017}. The technique has also been employed for computer tomography of a GDI spray showing some axial asymmetry of the liquid mass distribution.\cite{Cai2003}.
\\
Despite those noticeable advantages, synchrotron sources have some limitations when it comes to spray imaging: 
1) They usually have very small beam divergence ($\ll 1$\,mrad) requiring long beam transport lines, resulting in imaged areas of only a few mm.
2) In contrast to soft X-rays, the hard X-rays used in previous works have low absorption through the injected liquid. To increase absorption, a contrast agent such as potassium iodide, KI, is often added up to a non-negligible concentration for single-shot imaging. Such additives changes the liquid viscosity as well as surface tension (data given in \href{link}{Supplement 1}), thus, directly impacting the atomization process.  
3) Synchrotron sources have limited availability and high running costs.
\begin{figure*}[th]
\centering
\fbox{\includegraphics[width=179mm,scale=1]{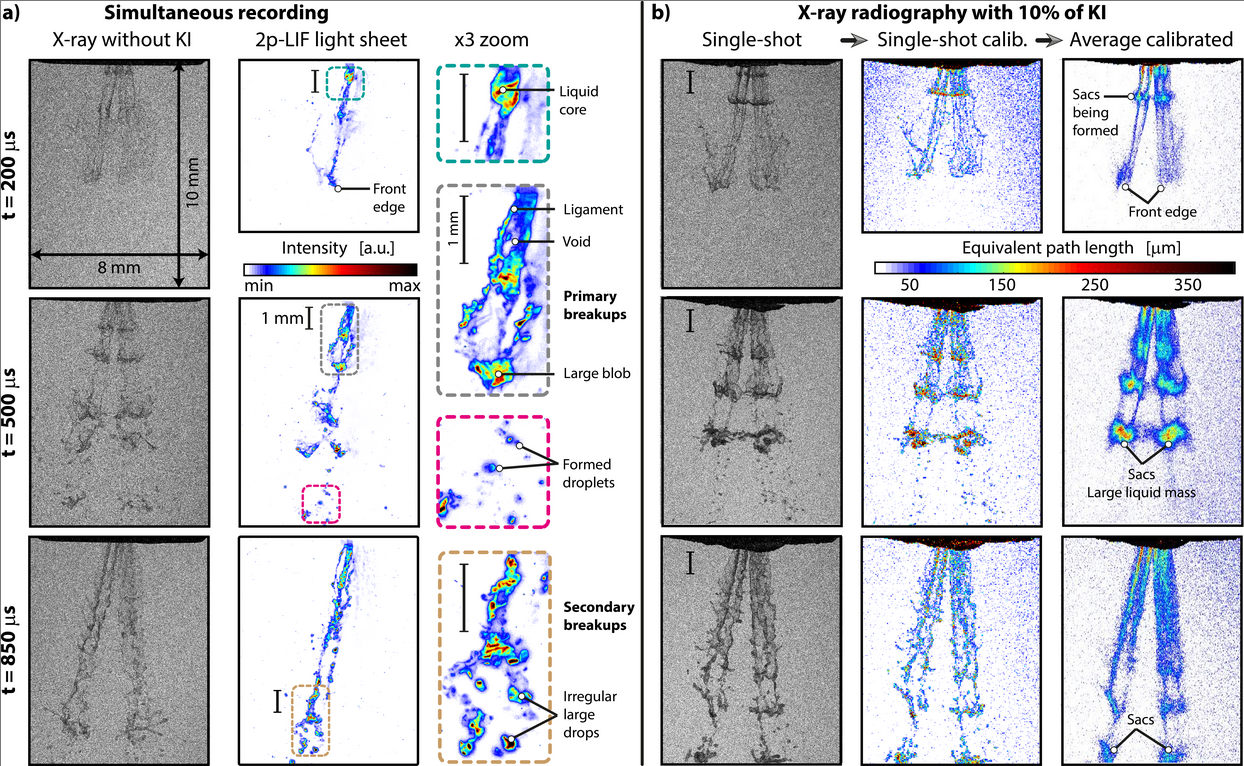}}
\caption{ Image results and comparison. In (a) simultaneous recordings of X-ray radiography and planar 2p-LIF are shown. In this case no KI was added to the injected water. In (b) the injected water contains 10\% of KI. In this case X-rays absorption is significant, increasing image contrast. By averaging and calibrating 50 single-shot images, the liquid equivalent liquid path length is extracted.}
\label{fig:Comparison_X_ray_fluo}
\end{figure*}
\\
Here we demonstrate the unique possibility of simultaneously using soft X-ray radiography and planar 2p-LIF to image the disintegration of liquid jets. This challenging configuration was made possible by means of a laser plasma accelerator depicted in Fig.\ref{fig:setup}. The concept was introduced for the first time in 1979 \cite{Tajima1979}. In this scheme, an ultrashort laser pulse reaching intensities above $10^{18}\,\text{W}/\text{cm}^2$ is used to ionize a gas medium, thus producing a plasma and exciting a plasma wave as shown in the inset of Figure \ref{fig:setup}. If the focal spot size and pulse duration match the plasma wavelength, the plasma wave becomes highly non-linear and a cavity partially depleted of electrons forms behind the laser pulse \cite{Lu2007}. In this cavity, there are strong focusing and accelerating electric fields (up to 100’s of GV/m) which allow accelerating co-propagating electrons up to, energies around hundreds of MeV and few mrad divergence, over a few millimeters. Several methods exist to inject electrons into this cavity. In this work and for the sake of stability we used a gas mixture (helium and nitrogen) to inject electrons via the ionization injection mechanism \cite{Pak2010}. It utilizes the fact that the most inner shells in nitrogen are ionized only at the peak of the laser pulse to release electrons directly inside the cavity.
\\
The laser system from the Lund High-Power Laser Facility is used in this work. The system provides 800 mJ, 38\,fs laser pulses which are focused down to 13\,µm reaching a peak intensity of $10^{19}\,\text{W}/\text{cm}^2$. At focus, the laser crosses a 1.5\, mm gas jet made of a 99\% He, 1\% N$_2$ mixture of density $\sim 1\times10^{19}\,\text{cm}^{-3}$. This results in the production of an electron bunch with 10's of pC, 5\,mrad divergence and energy up to 200\,MeV.
\\
During acceleration, the electrons oscillate transversely inside the plasma cavity; this motion leads to the generation of keV X-rays in the forward direction \cite{Rousse2004}. The X-rays are emitted with a characteristic spectrum ranging from 1 up to 20\,keV that typically peaks around a few keV, a maximum of $4\times10^8$  photons, a source size of less than 3\,µm, a divergence of less than 30\,mrad and an estimated duration of $\approx$ 10\,fs according to numerical simulations. During transport outside of the vacuum chamber, the various filters (Al, Kapton, and Beryllium windows) absorb the low energy photons thus shifting up the spectral maximum to $5.5$\,keV, with $5\times10^7$ photons detected by the camera (data given in \href{link}{Supplement 1}). 
\\
This radiation source has been used for various experiments \cite{Albert2016} ranging from imaging of static objects \cite{Svendsen2018} to spectroscopy of dynamical processes \cite{mahieu2018} and large efforts are being made to improve its quality \cite{Doepp2018}. For various reasons, it is very well adapted to imaging transient phenomena such as liquid breakups occuring in sprays atomization. 1) The energy range of the X-rays (1- 10 keV) is ideal for measuring absorption in 100\,µm of liquid. 2) As the approach is based on using high energy femtosecond laser pulses a portion of it can easily be used to simultaneously image the spray using 2p-LIF at no extra cost. 3) The divergence of the ultra-short X-ray pulses can result in a beam of relatively large diameter, in the cm range after 1\,m of propagation.
\\
The spray system used here is produced by a commercially available fuel port injector, Bosch EV1 4-holes nozzle, with orifice size of 280µm, running at 4.5 bar liquid injection pressure. The injected liquid is either composed, here, of water + fluorescein only or of water + fluorescein + 10\% KI (potassium iodide) in order to increase X-ray absorption.
Note that this mass concentration of KI changes the surface tension of the liquid by 1.1\% and the viscosity by 12\%, thus slightly impacting the liquid breakup. On the contrary, the fluorescein dye was added at only 0.1\%, resulting in negligible effects on the liquid properties (see details in \href{link}{Supplement 1}). 
After crossing the spray, the X-ray beam is recorded using a 4 Megapixel X-ray camera (Andor iKon-L SO CCD). The resolution is limited by the pixel size of 13.5\,µm, corresponding to 11.3\,µm for the spray due to the magnification from the beam divergence. The fluorescein dye is added in the injected liquid in order to generate a two-photon fluorescence signal in the range 500-600 nm \cite{Berrocal2019}. The two-photon absorption process is induced by extracting a small fraction of the incident high energy beam $\sim$ 10 mJ, it is redirected towards the spray and focused into a light sheet with a cylindrical lens. A 5.5 Megapixel s-CMOS camera (Andor Zyla) is placed vertically to record the fluorescence signal.  The camera was used with a Micro-Nikkor lens at F\#4 and each pixel was resolving an area of 8\,µm x 8\,µm. The details of the experiment are provided in Fig.\ref{fig:setup}.
\\
Figure \ref{fig:Comparison_X_ray_fluo}(a) shows single-shot X-ray (on the left) and light sheet 2p-LIF (on the right) images recorded simultaneously. These results correspond to three different times after the visible start of injection: 200\,µs, 500\,µs and 850\,µs respectively. The global features of the jet are visible from the X-ray images and can be quantified, but the noise is too high to distinguish the fine details such as individual droplets. On the contrary, the 2p-LIF image allows a clear visualization of individual droplets, liquid blobs and ligaments. The high contrast obtained from 2p-LIF detection is due to the reduced amount of fluorescence signal originating from multiple light scattering. In addition, the light sheet configuration allows to optically section the spray. This provides spray details which are not accessible with line-of-sight configurations.
Figure \ref{fig:Comparison_X_ray_fluo}(a) demonstrates the possibility of using laser driven X-rays to image jets of small equivalent path length (EPL) without the need of absorbing additives. In order to improve the image contrast from the X-ray images, a moderate amount of KI is added in the injected water as a contrast agent. The resulting images are shown in \ref{fig:Comparison_X_ray_fluo}(b), for non-calibrated single-shots (left side), calibrated single-shots (center) and images averaged over 50 single-shots (right side).  Those results show how the liquid mass is statistically distributed in space and how the jet is evolving over time. Note that the EPL measured right at the nozzle exit is $\approx$ 250\,µm corresponding well to the size of each orifice. The calibration uses X-ray transmission tables for water and KI \cite{@henke}. The sensitivity of the equivalent path length on a single-shot image is also deduced by evaluating the amount of liquid that is necessary to generate a signal higher than the surrounding noise. This sensitivity equals $60$\,µm of pure water and $25$\,µm of water + 10\% KI (see \href{link}{Supplement 1} for detailed calculation).

To conclude, we have shown the possibility of utilizing the intense femtosecond laser pulse used in laser plasma accelerator for simultaneous X-ray absorption and 2p-LIF imaging of a spray system typically used in internal combustion engines. The combination of advanced optical and X-ray techniques proposed here, provides complementary and unique descriptions of the probed spray. In addition, the measurement sensitivity of the equivalent path length from single-shot images - $25$\,µm for 10\% KI in water - is found to be higher than what has been achieved so far with synchrotron \cite{Halls2017}. Future strategic modifications of the presented setup such as reducing the distance between the source and the spray or using thinner windows and foils will further improve the measurement sensitivity, so that no contrast agent will be needed. Also, by accurately rotating the injector, three-dimensional reconstruction of the liquid mass distribution through the entire spray can be obtained. In addition, velocity vectors or liquid temperature can also be obtained from 2p-LIF measurements. Finally, this proof-of-concept article demonstrates the first use of laser driven X-rays for imaging an atomizing jet, paving the way for the future characterization of a wide range of spray systems generated from different nozzle geometry and running at more challenging operating conditions, such as at high liquid injection pressure.

\section*{Funding Information}
European Research Council (ERC): "Spray-Imaging" - 638546; Vetenskapsrådet (2016-03894 and 2015-03749); Knut and Alice Wallenberg Foundation. 

\section*{Disclosures}
The authors declare no conflicts of interest.

\bigskip \noindent See \href{link}{Supplement 1} for supporting content.

\bibliography{sample}

\begin{thebibliography}{10}
\newcommand{\enquote}[1]{``#1''}

\bibitem{Paciaroni2004}
M.~Paciaroni and M.~Linne, {\protect\JournalTitle{Applied Optics}} \textbf{43},
  5100 (2004).

\bibitem{Linne2009}
M.~A. Linne, M.~Paciaroni, E.~Berrocal, and D.~Sedarsky,
  {\protect\JournalTitle{Proceedings of the Combustion Institute}}
  \textbf{32(2)}, 2147–2161 (2009).

\bibitem{Idlahcen2012}
S.~Idlahcen, C.~Roz{\'e}, L.~M{\'e}{\`e}s, T.~Girasole, and J.-B. Blaisot,
  {\protect\JournalTitle{Experiments in Fluids}} \textbf{52}, 289 (2012).

\bibitem{Rahm2016}
M.~Rahm, Z.~Falgout, D.~Sedarsky, and M.~Linne, {\protect\JournalTitle{Optics
  express}} \textbf{24}, 4610 (2016).

\bibitem{Berrocal2008}
E.~Berrocal, E.~Kristensson, M.~Richter, M.~Linne, and M.~Ald\'en,
  {\protect\JournalTitle{Optics Express}} \textbf{16(22)}, 17870–17881
  (2008).

\bibitem{Kristensson2018}
E.~Kristensson and E.~Berrocal, {\protect\JournalTitle{Scientific Report}}
  \textbf{8(1)}, 11751 (2018).

\bibitem{Berrocal2012}
E.~Berrocal, J.~Johnsson, E.~Kristensson, and M.~Ald\'en,
  {\protect\JournalTitle{Journal of the European Optical Society Rapid
  Publication}} \textbf{8(1)}, 12015 (2012).

\bibitem{Berrocal2016}
E.~Berrocal, S.-G. Pettersson, and E.~Kristensson,
  {\protect\JournalTitle{Optics Letter}} \textbf{41}, 5612 (2016).

\bibitem{Kristensson2012}
E.~Kristensson, E.~Berrocal, and M.~Ald\'en, {\protect\JournalTitle{Optics
  Express}} \textbf{20}, 14437 (2012).

\bibitem{Berrocal2019}
E.~Berrocal, J.~P. C.~Conrad, C.~L. Arnold, M.~Wensing, M.~Linne, and
  M.~Miranda, {\protect\JournalTitle{OSA Continuum}} \textbf{2}, 983 (2019).

\bibitem{Heindel2019}
T.~J. Heindel, {\protect\JournalTitle{Atomization in sprays}} \textbf{28}, 1029
  (2018).

\bibitem{Mishra2017}
Y.~N. Mishra, E.~Kristensson, M.~Koegl, J.~J\"onsson, L.~Zigan, and
  E.~Berrocal, {\protect\JournalTitle{Experiments in Fluids}} \textbf{58(9)},
  110 (2017).

\bibitem{Linne2012}
M.~A. Linne, {\protect\JournalTitle{Experiments in fluids}} \textbf{53}, 655
  (2012).

\bibitem{Powell2000}
C.~F. Powell, Y.~Yue, R.~Poola, and J.~Wang, {\protect\JournalTitle{Journal of
  Synchrotron Radiation}} \textbf{7} (2000).

\bibitem{MacPhee2002}
A.~G. MacPhee, M.~W. Tate, C.~F. Powell, M.~J.~R. Y.~Yue., A.~Ercan,
  S.~Narayanan, E.~Fontes, J.~Walther, J.~Schaller, S.~M. Gruner, and J.~Wang,
  {\protect\JournalTitle{Science}} \textbf{295}, 1261 (2002).

\bibitem{Halls2017}
B.~R. Halls, C.~D. Radke, B.~J. Reuter, A.~L. Kastengren, J.~R. Gord, and T.~R.
  Meyer, {\protect\JournalTitle{Optics Express}} \textbf{25}, 1605 (2017).

\bibitem{Cai2003}
W.~Cai, C.~F. Powell, Y.~Yue, S.~Narayanan, J.~Wang, M.~W. Tate, M.~J. Renzi,
  A.~Ercan, E.~Fontes, and S.~M. Gruner, {\protect\JournalTitle{Applied Physics
  Letter}} \textbf{83}, 1671 (2003).

\bibitem{Tajima1979}
T.~Tajima and J.~M. Dawson, {\protect\JournalTitle{Physical Review Letters}}
  \textbf{43}, 267 (1979).

\bibitem{Lu2007}
W.~Lu, M.~Tzoufras, C.~Joshi, F.~S. Tsung, W.~B. Mori, J.~Vieira, R.~A.
  Fonseca, and L.~O. Silva, {\protect\JournalTitle{Phys. Rev. ST Accel. Beams}}
  \textbf{10} (2007).

\bibitem{Pak2010}
A.~Pak, K.~A. Marsh, S.~F. Martins, W.~Lu, W.~B. Mori, and Joshi,
  {\protect\JournalTitle{Phys. Rev. Lett.}} \textbf{104} (2010).

\bibitem{Rousse2004}
A.~Rousse, K.~T. Phuoc, R.~Shah, A.~Pukhov, E.~Lefebvre, V.~Malka, S.~Kiselev,
  F.~Burgy, J.~Rousseau, D.~Umstadter, and D.~Hulin,
  {\protect\JournalTitle{Phys. Rev. Lett.}} \textbf{93}, 135005 (2004).

\bibitem{Albert2016}
F.~Albert and A.~G.~R. Thomas, {\protect\JournalTitle{Plasma Physics and
  Controlled Fusion}} \textbf{58} (2016).

\bibitem{Svendsen2018}
K.~Svendsen, I.~G. Gonz\'{a}lez, M.~Hansson, J.~B. Svensson, H.~Ekerfelt,
  A.~Persson, and O.~Lundh, {\protect\JournalTitle{Optics express}}
  \textbf{26}, 33930 (2018).

\bibitem{mahieu2018}
B.~Mahieu, N.~Jourdain, K.~T. Phuoc, F.~Dorchies, J.-P. Goddet, A.~Lifschitz,
  P.~Renaudin, and L.~Lecherbourg, {\protect\JournalTitle{Nature
  Communications}} \textbf{9} (2018).

\bibitem{Doepp2018}
A.~D\"{o}pp, L.~Hehn, J.~G\"{o}tzfried, J.~Wenz, M.~Gilljohann, H.~Ding,
  S.~Schindler, F.~Pfeiffer, and S.~Karsch, {\protect\JournalTitle{Plasma
  Physics and Controlled Fusion}} \textbf{5}, 199 (2018).

\bibitem{@henke}
\enquote{X-ray interactions with matter,}
  \url{http://henke.lbl.gov/optical_constants/}.

\end{thebibliography}

\bibliographyfullrefs{sample}


\end{document}